\documentclass[12pt]{iopart}
  \expandafter\let\csname equation*\endcsname\relax
  \expandafter\let\csname endequation*\endcsname\relax
\usepackage{amsmath, amssymb}

\begin{document}

\title{A Master Equation for Gravitational Decoherence: Probing the Textures of Spacetime}

\author{C. Anastopoulos$^1$ and B. L. Hu$^2$}

\address{$^1$Department of Physics, University of Patras, 26500 Patras, Greece.}

\address{$^2$Maryland Center for Fundamental Physics and Joint Quantum Institute,\\ University of
Maryland, College Park, Maryland 20742-4111 U.S.A.}

\ead{anastop@physics.upatras.gr,blhu@umd.edu}

\date{July 5, 2013}

\begin{abstract}
We give a first principles derivation  of  a master equation for the evolution of a quantum matter field in a linearly perturbed Minkowski spacetime, based solely on quantum field theory and general relativity. We make no additional assumptions nor introduce extra ingredients, as is often done in alternative quantum theories. When the quantum matter field is projected to a one-particle state, the master equation for a non-relativistic quantum particle in a weak gravitational field predicts decoherence in the \textit{energy basis}, in contrast to most existing theories of gravitational decoherence.  We point out the gauge nature of time and space reparameterizations in matter-gravity couplings, and warn that `intrinsic' decoherence or alternative quantum theories invoking stochastic dynamics arising from temporal or spatial fluctuations violate this fundamental symmetry of classical general relativity. Interestingly we find that the decoherence rate depends on \textit{extra parameters} other than the Planck scale, an important feature of gravitational decoherence.  This is similar to the  dependence of the decoherence rate of a quantum Brownian particle to the temperature and spectral density of the environment  it interacts with. The corresponding features when gravity acts as an environment in decohering quantum objects are what we call the \textit{`textures' of spacetime}. We point out the marked difference between the case when gravity is represented as a background spacetime versus the case when gravity acts like a thermodynamic bath to quantum particles. This points to the possibility of using gravitational decoherence measurements to discern whether gravity is intrinsically elemental or emergent.
\end{abstract}

\section{Introduction}

Since gravitational decoherence conveys different meanings in different contexts for the purpose of clarification we begin with a qualifying description of what quantum, intrinsic and gravitational each refers to in relation to decoherence.

\subsection{Quantum, Intrinsic and Gravitational Decoherence}

\textbf{Quantum decoherence} refers to the loss of coherence in a quantum system for
various reasons, commonly due to its interaction with an environment.
It could be formulated in different ways, through decoherent or
consistent histories \cite{conhis}  or via open-system dynamics
\cite{envdec,Dav,BP}. Much work has been done since the beginning of the
90's that we now think we have a better understanding of this issue
(see books and reviews, e.g., in \cite{decrev}). The source and means which cause a quantum system to decohere may come from ordinary matter, quantum fields, or gravitational fields. Here then rests further distinction between the so-called `intrinsic' or `fundamental' decoherence on the one hand and gravitational decoherence on the other. We give a brief sketch of their differences, as these terms are used by different
authors in special contexts for specific purposes.
With this chart we can then spell out what we intend to investigate in this paper.

\textbf{Intrinsic or fundamental decoherence} refers to some
intrinsic or fundamental conditions or processes which engender
decoherence in varying degrees but universal to all quantum systems.
This could come from (or could account for) the uncertainty relation
\cite{WignerUR, Karol}, some fundamental imprecision in the measuring
devices (starting with clocks and rulers) \cite{GPP}, in the
dynamics \cite{MilIntDec}, or in treating time as a statistical variable
\cite{Bonifacio}. Hereby lies the possibility of alternative theories
of quantum mechanics, such as theories based on a stochastic
Schr\"odinger equation. This alternative form has been proposed by many
authors, notably, Ghirardi, Rimini, Weber and Pearle \cite{GRWP}
Diosi \cite{Diosi},  Gisin \cite{Gisin}, Penrose \cite{Penrose},
Persival \cite{Persival}, Hughston \cite{Hughston} et al who
suggest gravity as the origin of the modification. (See \cite{Diosi05} for a comparison of different
approaches.)  Here, the search is
mainly motivated by an uneasiness in the awkward union between
general relativity and quantum mechanics, and the context can be in
the settings of quantum gravity or at today's low energy with a
classical spacetime structure and quantum matter fields.  Adler
\cite{Adler} views quantum mechanics as emergent from a more
fundamental theory while 't Hooft views quantum mechanics as a clever way of bookkeeping classical events \cite{tHooft}. For a recent discussion of the differences between intrinsic and environment-induced decoherence, see Ref. \cite{Stamp}.

In an earlier paper \cite{IntDec}, we explored a range of related issues,
including the meaning of modified uncertainty relations, the
interpretations of the Planck scale, the distinction between quantum
and stochastic fluctuations and the role of the time variable in
quantum mechanics. We examined the specific physical assumptions that
enter into different approaches to the subject, in particular, the modeling of  space and time fluctuations by stochastic processes.

Some protagonists believe that gravity may be needed to make quantum
physics more complete, and since universal conditions are involved in
these investigations, gravitational decoherence is often brought up
for this purpose.

\textbf{Gravitational decoherence} refers to the effect of gravity on
the decoherence of quantum systems \cite{Diosi, Penrose, Persival, Kay, WBM06, Breu, Diosi2, Blenc}. In principle, it also pertains to
quantum gravitational effects  operative presumably at the Planck
scale, but we separate our consideration of these two regimes so the
role of gravity will not be confused.  We save the term gravitational
decoherence to refer to gravity as described by general relativity.
For this,  even weak gravity is thought to act differently in
bringing decoherence to a quantum system than other matter fields.
For a discussion on issues of decoherence in
quantum gravity, see, e.g., \cite{AHdecQG} and references therein.

\subsection{Issues, Setup and Methodology}

{\em Issues.} With the meaning defined above, the questions we ask in this paper are.
\begin{enumerate}
\item Whether and how quantum matter is decohered by a gravitational field, allowing for weak-field conditions only.
\item How gravitational decoherence differs from decoherence by a non-gravitational environment.
\item What are the special features of gravitational decoherence, in particular, the decoherence rate and the associated basis.
\end{enumerate}

We find answers to these questions from a first-principles treatment of gravitational decoherence, viewing gravitational perturbations as an environment that affects the evolution of quantum particles. We employ general relativity (GR) for the description of gravity and quantum field theory (QFT) for the description of the matter degrees of freedom. Our derivations proceed from the general case to specific systems. We want to establish a general method for the study of gravitational decoherence that can be applied to many different physical situations. We try to avoid complex modeling assumptions, having as our main guide the mathematical structures of the two well-proven theories.

It is our hope that the results we obtain here, including a general master equation for the quantum matter field may serve as a lantern for illuminating the proposed alternative theories of quantum physics and/or gravity mentioned above. Our approach allows anyone to identify what and where new ingredients are added as assumptions,  how these assumptions may be at odds with  the principles of general relativity and quantum mechanics and whether and where these theories are imbued with intrinsic contradictions with GR and QFT.   By laying out explicitly the logical implications of GR and QFT for gravitational decoherence, we want to provide a standard with which the protagonists of all new theories can compare, explain and better justify the {\it raison d'\^etre} of their creation.

\bigskip

\noindent {\em Setup.} The system under consideration is a massive scalar field, interacting with a gravitational field as its environment.
Gravity is described by classical general relativity. In the weak field limit, we describe gravitational perturbations in the linearized approximation. In the spirit of decoherence studies via the theory of open quantum systems, we trace out the gravitational degrees of freedom and derive a master equation for the quantum matter degrees of freedom. Since we describe the matter degrees of freedom by a quantum field, the resulting master equation can be applied to the analysis of gravitational decoherence in systems with an arbitrary number of  particles. To obtain a master equation for a single particle we project the quantum field master equations to the one-particle subspace.

The weak-field assumption for the gravitational field is essential to our method. The tracing-out of the gravitational field is not technically feasible, unless we specialize to the case of linearized gravitational perturbations. With this assumption, the system under consideration is formally similar to a quantum Brownian motion (QBM) model \cite{QBM}. The weak-field regime is in principle testable by experiments at low energy  (in contrast to strong field conditions, as found in the early universe or late time black holes, which are intellectually challenging issues, yet hitherto unreachable.).

The master equation we derive here applies to any type of quantum matter fields in addition to scalar, any number of particles, in the ultra-relativistic as well as the non-relativistic regimes. It can serve as a basis for exploring gravitationally-induced effects in significantly more complex regimes than the one considered here.

\bigskip

\noindent {\em Methodology.} Our starting point is the classical action for gravity interacting with a scalar field. We linearize the Einstein-Hilbert action around Minkowski spacetime, perform a 3+1 decomposition of the action and construct the associated Hamiltonian. We identify the constraints of the system and solve them at the classical level, expressing the Hamiltonian in term of the true physical degrees of freedom of the theory, namely, the transverse-traceless perturbations for gravity and the scalar field.

We then quantize the scalar field and the gravitational perturbations and trace-out the contribution of the latter. A key input in this stage is a specification of an initial state for the gravitational perturbations. We consider an initial condition that interpolates between the regime of negligible (vacuum) perturbations and strong classicalized perturbations. The initial state is defined in terms of a free parameter $\Theta$ that can be loosely interpreted as the noise temperature of the perturbations.  $\Theta$ may be viewed as a parameter which conveys coarse-grained information reflective of the micro-structures of spacetime, similar to temperature with regard to molecular motion.
 It is in this sense that we think gravitational decoherence may reveal the underlying `textures' of spacetime beneath that described by classical general relativity. We will have more to say about this aspect later.

After this we  follow the standard methodology of open quantum systems \cite{Dav, BP} in order to derive the 2nd order (perturbative) master equation for the quantum scalar field. This master equation applies to configurations with any number of particles. We project the master equation to the single-particle subspace and we derive a master equation for a single particle. The latter simplifies significantly in the non-relativistic regime, and allows for the determination of the decoherence rate.

\subsection{Main Results} 


\begin{enumerate}

\item We derive from accepted theories for quantum matter (QFT) and classical gravity (GR) a master equation describing a moving particle interacting with a weak gravitational field. A special feature of decoherence by the gravitational field (in the non-relativistic limit) is the decoherence in the energy (momentum squared) basis, but not (directly) to decoherence in the position basis.

\item  We examine the significance of space and time reparameterizations in the description of a quantum field interacting with linearized gravity. We find that in order to obtain a Poincar\'e covariant description of the quantum field on Minkowski spacetime, it is {\em necessary} to fix the gauge. A gauge-invariant treatment of the associated constraints  does not appear compatible with the structures of Poincar\'e covariant QFT. This is the physical rationale for Penrose's gravity-induced decoherence \cite{Penrose}. Our results lead to a  reformulation of his argument from a different perspective.

\item Many approaches to gravitational or fundamental decoherence proceed by modeling temporal or spatial fluctuations in terms of  stochastic processes. However, such fluctuations correspond to time or space reparameterizations, which are pure gauge variables, with no dynamical content, according to classical GR.
     The assignment of dynamical content to such reparameterizations implicitly presupposes an underlying theory that violates the fundamental symmetry of classical GR.

\item The time-scale of decoherence in our approach depends crucially on a new parameter $\Theta$, related to the strength of gravitational perturbations, which is not simply related to the Planck length. We argue that this result is generic in any treatment of decoherence from the perspective of open quantum systems.  The decoherence rate should depend not only on the matter-gravity coupling, but also on the intrinsic properties of the environment,  such as its spectral density which reflects to some extent the characteristics of its micro-physics composition. Measurement of the gravitational decoherence rate, if this effect due to gravity can be cleanly separated from other sources, may thus provide valuable information about the gravitational ``noise temperature" or ``spectral density", what we call the underlying ``textures" of spacetime.

\end{enumerate}

We now proceed with the derivations.

\section{Quantum Matter Scalar Field Interacting with Linearized Gravity}

Our starting point is the classical action for gravity interacting with a scalar field. We linearize the Einstein-Hilbert action around Minkowski spacetime. Then, we perform a 3+1 decomposition of the action and construct the associated Hamiltonian. We identify the constraints of the system and we solve them at the classical level. We conclude with an expression of the  Hamiltonian for the system in terms of the true physical degrees of freedom of the theory. This expression forms the basis for   quantization in the following section.

\subsection{The action}
The action for a classical scalar field theory describing the matter degrees of freedom $\phi$
interacting with the gravitational field is
\begin{eqnarray}
S[g,\phi] = \frac{1}{\kappa}\int  d^4x \sqrt{-g} R +  \int d^4x \sqrt{-g}
\left(-\frac{1}{2} g^{\mu \nu} \nabla_{\mu} \phi \nabla_{\nu}\phi -
\frac{1}{2}{ m^2} \phi^2\right), \label{Scov}
\end{eqnarray}
where $\nabla_{\mu}$ is the covariant derivative defined on a background spacetime with Lorentzian metric  $g_{\mu\nu}$, $R$ is the spacetime Ricci scalar and  $m$ is the scalar-field's mass.

We assume for the  spacetime manifold a spacelike foliation in the form $\Sigma \times R$ characterized by the
coordinate $t$ in the time direction and coordinates $x^i$ on a
spacelike surface $\Sigma$.  We denote the Riemannian metric on $\Sigma$ as $h_{ij}$ and the corresponding Ricci scalar as ${}^3R$. With this we perform a $3+1$ decomposition of the action in Eq. (\ref{Scov}) resulting in:
\begin{eqnarray}
S_{3+1}[h_{ij}, \phi, N, N^i] = \frac{1}{\kappa} \int dt d^3 x N \sqrt{h} \left[ K_{ij}K^{ij} -K^2 +{}^{(3)}R \right. \\ \nonumber
\left. + \frac{1}{2N^2}\dot{\phi}^2 - \frac{1}{2} (h^{ij} - \frac{N^iN^j}{N^2}) \nabla_i \phi \nabla_j \phi -\frac{1}{N^2} \dot{\phi} N^i\nabla_i\phi \right], \label{S3+1}
\end{eqnarray}
where $N$ is the lapse function, $N^i$ the shift vector, and
\begin{eqnarray}
K_{ij} = \frac{1}{2N} \left(\dot{h}_{ij} - \nabla_i N_j - \nabla_j N_i\right)
\end{eqnarray}
is the extrinsic curvature on $\Sigma$. The dot denotes taking the Lie
derivatives with respect to the vector field $\frac{\partial}{\partial t}$.

The standard description of linearized general relativity entails
expanding the four-metric around a flat (Minkowski) background,
keeping second-order terms to the perturbations in the gravitational
part of the action and first-order terms in the coupling to matter.
The approximation corresponds to an expansion in powers of $\kappa$, defined as follows.

We consider perturbations around the Minkowski spacetime ($N = 1, N^i
= 0, h_{ij} = \delta_{ij}$) that are first-order with respect to
$\kappa$. That is, we write
\begin{eqnarray}
h_{ij} = \delta_{ij} + \kappa \gamma_{ij}, \hspace{1cm} N = 1 +
\kappa n, \hspace{1cm} N^i = \kappa n^i, \label{expand}
 \end{eqnarray}
 and we keep in Eq. (\ref{S3+1}) only terms up to first order in $\kappa$. We obtain
\begin{eqnarray}
S_{lin}[\gamma_{ij},\phi, n, n^i] =  \int dt d^3 x  \left(\frac{1}{2} \dot{\phi}^2 - \frac{1}{2}  \partial^i \phi \partial_i \phi - \frac{1}{2} m^2 \phi^2 \right) \nonumber \\
+ \kappa \int dt d^3x \left[ \frac{1}{4} (\dot{\gamma}_{ij} -2\partial_{(i} n_{j)})( \dot{\gamma}^{ij} - 2 \partial^{(i} n^{j)}) - \frac{1}{4} (\dot{\gamma} - 2\partial_i n^i)^2 \right.
\nonumber \\
\left.
- V[(\partial \gamma)^2] +n(\partial_i\partial_j \gamma - \partial^2\gamma)\right]  \nonumber \\
+ \frac{\kappa}{2} \int dt d^3x\left[(\frac{1}{2}\gamma - n)
\dot{\phi}^2 - 2n^i \dot{\phi}\partial_i \phi  + \gamma^{ij}
\partial_i \phi \partial_j \phi - (n+ \frac{1}{2} \gamma)
(\partial^i\phi \partial_i \phi + m^2 \phi^2)\right]. \hspace{1cm} \label{slin}
\end{eqnarray}
The indices in Eq. (\ref{slin}) are raised and lowered with the background 3-metric $\delta_{ij}$. We have defined $\gamma = \delta^{ij} \gamma_{ij}$. The "potential" $V[(\partial \gamma)^2]$ corresponds to the second order terms in the expansion of $\sqrt{h} {}^3R$ with respect to $\gamma$. Explicitly,
\begin{eqnarray}
V = -\frac{1}{2} \partial_k\gamma_{ij} \partial^i \gamma^{kj} - \frac{1}{2} \partial_k \gamma \partial^k \gamma + \partial_i \gamma \partial_k \gamma^{ik} + \frac{1}{4} \partial_k \gamma_{ij} \partial^k \gamma^{ij}.
\end{eqnarray}

The first term in Eq. (\ref{slin}) is the action for a free scalar
field on Minkowski spacetime, the second term describes the self-dynamics
of the perturbations and the third term describes the
matter-gravity coupling. Note that the terms for the gravitational self-dynamics
and the matter-gravity coupling are of the same order in $\kappa$.

\subsection{The Hamiltonian}

To obtain the Hamiltonian we perform the Legendre transform of the
Lagrangian density ${\cal L}_{lin}$ associated  to the action Eq.
(\ref{slin}). The conjugate momenta $\Pi^{ij}$ and $\pi$ of
$\gamma_{ij}$ and $\phi$ respectively are
\begin{eqnarray}
\Pi^{ij} &:=& \frac{\partial {\cal L}_{lin}}{\partial \dot{\gamma}_{ij}} = \frac{\kappa}{2} \left(\dot{\gamma}_{ij} -  \dot{\gamma}\delta^{ij} + \partial^in^j + \partial^j n^i -2 \partial_k n^k \delta^{ij}\right),\\
\pi &:=&  \frac{\partial {\cal L}_{lin}}{\partial \dot{\phi}} =
\dot{\phi} + \kappa \left[(\frac{1}{2} \gamma - n) \dot{\phi} - n^i
\partial_i \phi\right].
\end{eqnarray}
The conjugate momenta $\Pi_{n} = \partial {\cal
L}_{lin}/\partial{\dot{n}}$ and $\Pi^i_{
\overrightarrow{n}} = \partial {\cal L}_{lin}/\partial{\dot{n_i}}$
vanish identically. Thus, the equations $\Pi_{n} = 0$ and
$\Pi_{ \overrightarrow{n}}^i = 0$ define primary
constraints.

The Hamiltonian $H = \int d^3x (\Pi^{ij} \dot{\gamma}_{ij} + \pi \dot{\phi}
- {\cal L}_{lin})$  is
\begin{eqnarray}
H =  \int d^3 x \left[\left(\frac{\Pi^{ij}\Pi_{ij} - \frac{1}{2}
\Pi^2}{\kappa} + \kappa V[(\partial \gamma)^2]\right) +
\mathfrak{e}(\phi, \pi) \right. \nonumber \\
\left.
-\frac{\kappa}{2}   \left[\gamma \mathfrak{e}(\phi, \pi) +  \gamma^{ij} \partial_i\phi \partial_j \phi - \gamma (\partial_k \phi \partial^k \phi +m^2 \phi^2)\right] \right. \nonumber \\
+ \left.   n \left[\partial^2\gamma - \partial_i \partial_j
\gamma^{ij} + \mathfrak{e}(\phi,\pi)\right] + n_i \left[-2
\partial_j \Pi^{ji} +  \kappa \mathfrak{p}^i(\Pi, \phi) \right] \right],
\label{hlin}
\end{eqnarray}
where $\Pi = \Pi^{ij} \delta_{ij}$, and
\begin{eqnarray}
\mathfrak{e}(\phi, \Pi) = \frac{1}{2} \pi^2 + \frac{1}{2}
\partial_i\phi
\partial^i \phi + \frac{1}{2} m^2 \phi^2
\end{eqnarray}
is the energy density of the scalar field, and
\begin{eqnarray}
\mathfrak{p}^i(\phi,\pi) = \pi \partial^i \phi
\end{eqnarray}
is the momentum density (energy flux).

Eq. (\ref{hlin}) can also be obtained from the full gravitational Hamiltonian
\begin{eqnarray}
H = \int d^3x \left[ N \left(\frac{\Pi^{ij}\Pi_{ij} - \frac{1}{2}
\Pi^2}{\kappa \sqrt{h}} - \sqrt{h}{}^3R +
\mathfrak{h}(\phi,\pi,h_{ij})\right) \right. \nonumber \\
\left.
+ N^i \left(-2 \nabla_j
\Pi^j{}_i +  \mathfrak{h}_i(\phi, \pi ,h_{ij})\right) \right],
\label{hfull}
\end{eqnarray}
by expanding the metric variables around flat spacetime as in Eq. (\ref{expand}) and keeping terms  to first order in $\kappa$. Eq. (\ref{hfull}) applies to a larger class of field theories than the one we consider in this paper: any diffeomorphism-invariant action where matter fields do not couple to derivatives of the spacetime metric gives rise to a Hamiltonian of the form (\ref{hfull}) (plus additional constraints reflecting other gauge symmetries).

\subsection{Constraints  and Symmetries}

Eq. (\ref{hlin}) reveals the presence of secondary, first-class
constraints that arise from the usual scalar and vector constraints
of general relativity after linearization. The scalar constraint
${\cal C} = \partial^2\gamma - \partial_i \partial_j \gamma^{ij} +
\mathfrak{e} = 0$ generates the gauge transformations
\begin{eqnarray}
\delta \gamma_{ij} = 0, \hspace{0.5cm} \delta \Pi^{ij} = -
\partial^2\lambda \delta^{ij} + \partial^i \partial^{j}\lambda,
\hspace{0.5cm} \delta \phi = \lambda \frac{\delta H_0}{\delta \pi},
\hspace{0.5cm} \delta \pi = - \lambda  \frac{\delta H_0}{\delta \pi},
\label{gauge1}
\end{eqnarray}
where $H_0 = \int d^3x \mathfrak{e}$ is the field Hamiltonian at
Minkowski spacetime, and $\lambda$ is a scalar function on $\Sigma$. The
vector constraint ${\cal C}^i := -2 \partial_j \Pi^{ji} +  \kappa
\mathfrak{e}^i = 0$ generates the gauge transformations
\begin{eqnarray}
\delta \gamma_{ij} = \partial_i \lambda_j + \partial_j \lambda_i,
\hspace{0.5cm} \delta \Pi^{ij} =0, \hspace{0.5cm} \delta \phi =
\kappa \lambda^i \partial_i \phi, \hspace{0.5cm} \delta \pi = \kappa
\partial_i(\lambda^i \pi) \label{gauge2}
\end{eqnarray}
where $\lambda^i$ is a vector-valued function on $\Sigma$.

The gauge transformations Eqs. (\ref{gauge1}---\ref{gauge2}) correspond to temporal and spatial reparameterizations of the free fields. To see this, we write the longitudinal part of the metric perturbation
\begin{eqnarray}
{}^L\gamma_{ij}(x) = -i\int \frac{d^3k}{(2 \pi)^3} e^{-i {\bf k}
\cdot {\bf x}} [k_i q_j({\bf k}) + k_j q_i({\bf k})].
\end{eqnarray}
Then we define the function $ q_i(x) = \int \frac{d^3k}{(2
\pi)^3}e^{-i {\bf k}\cdot {\bf x}}q_i({\bf k})$, which  transforms
under Eq. (\ref{gauge2}) as $\delta q_i = \lambda_i$. It follows that
the scalar function $\tilde{\phi}(x) := \phi(x^i - q^i)$ and the scalar
density $\tilde{\pi}(x) :=  \pi (x^i -  q^i) - \partial_k q^k$ are invariant
under the transformations Eq. (\ref{gauge2}), and thus the true physical degrees
of freedom of the theory.

We decompose the Fourier transform of the conjugate momentum $\Pi^{ij}$ as
\begin{eqnarray}
\Pi^{ij}_{\bf k} = {}^T\Pi_{\bf k} (\delta_{ij} - \frac{k_i k_j}{k^2}) + {}^L \Pi^{ij}_{\bf k} + \bar{\Pi}^{ij}_{\bf k},
\end{eqnarray}
where ${}^L\Pi^{ij}_{\bf k}$ is the longitudinal component and $\bar{\Pi}^{ij}_{\bf k}$ is the transverse-traceless (TT) component. Then, the action of the scalar constraint corresponds to the transformation
\begin{eqnarray}
\delta {}^T\Pi_{\bf k} = k^2 \lambda_{\bf k}.
\end{eqnarray}
Thus, the variable $\tau_{\bf k} = {}^T\Pi_{\bf k} /k^2$ transforms as $\delta \tau_{\bf k} = \lambda_{\bf k}$.

Let
$(\phi(x,t), \pi(x,t))$ be solutions to the evolution equations
generated by the Hamiltonian $H_0 = \int d^3x \mathfrak{e}$  corresponding to the free field. Then the history $[\phi(x, t -
 \tau), \pi(x, t- \tau)]$ remains invariant under the
transformation (\ref{gauge1}).

It follows that the longitudinal part of the metric perturbation
${}^L\gamma_{ij}$ and the transverse trace ${}^T\Pi$ of the gravitational conjugate
momentum are pure gauge, reflecting the  freedom of space and time
reparameterization in the evolution of the matter degrees of freedom. The associated symmetry is {\em not} that of spacetime diffeomorphisms, but of the spacetime diffeomorphisms that preserve the spacelike foliation introduced for the purpose of the 3+1 decomposition.   The fact that time and space reparameterizations are not dynamical in general relativity is a very important criterion for all proposers of alternative models of gravitational decoherence to take notice. {\em Any postulate of dynamical or stochastic fluctuations that correspond to space and time reparameterizations conflicts with the fundamental symmetries of general relativity}.

\subsection{Gauge fixing}

We have implemented a 3+1 decomposition  using a foliation that corresponds to a Lorentz frame in the background Minkowski spacetime. However, the notion of a Lorentz frame does not remain invariant under the space and time reparameterizations induced by the constraints. This is problematic for the quantization of the matter degrees of freedom, because a quantum field theory in Minkowski spacetime is defined only with respect to a Lorentz frame, so that it carries a representation of the Poincar\'e group. Hence, for the purpose of quantization it is necessary to impose a gauge condition that preserves the Lorentz frame introduced by the foliation. Thus, we must assume that
 $q^i = 0 $ and $\tau =
0 $, or equivalently ${}^L\gamma_{ij} = 0 $ and ${}^T\Pi = 0$.

In this gauge, the scalar constraint becomes the Poisson equation
$\partial^2\gamma = - \mathfrak{e}$, which we   solve for $\gamma$ to
obtain
\begin{eqnarray}
\gamma(x) = - \int d^3x' \frac{\mathfrak{e}(x')}{4\pi |{\bf x} - {\bf
x'}|}. \label{gamma}
\end{eqnarray}
We also solve the vector constraint, in order to determine the
longitudinal part of $\Pi^{ij}$. We find
\begin{eqnarray}
{}^L\Pi^{ij}(x) = i \int \frac{d^3k}{(2 \pi)^3} e^{-i {\bf k} \cdot
{\bf x}} [k_i \nu_j({\bf k}) + k_j\nu_i({\bf k})], \label{pil}
\end{eqnarray}
where
\begin{eqnarray}
\nu_i({\bf k}) = \frac{\kappa}{2} \left(\delta_{ij} -
\frac{k_ik_j}{2k^2}\right) \tilde{\mathfrak{e}}^j({\bf k}).
\label{nu}
\end{eqnarray}
In Eq. (\ref{nu}),  $\tilde{\mathfrak{e}}^i({\bf k})$ denotes the
Fourier transform of the momentum density $\mathfrak{e}^i$.

Thus the true physical degrees of freedom in the system correspond to the
transverse traceless components $\bar{\gamma}_{ij}$,
$\bar{\Pi}^{ij}$ of the metric perturbations and conjugate momenta,
and to the matter variables $\phi$ and $\pi$. The Hamiltonian
(\ref{hlin}) then becomes
\begin{eqnarray}
H = \int d^3x  \left(\frac{1}{\kappa} \tilde{\Pi}^{ij}\, \bar{\Pi}_{ij}
+ \frac{\kappa}{4} \partial_k \bar{\gamma}_{ij} \partial^k \bar{\gamma}^{ij} + \mathfrak{e} - \frac{\kappa}{2} \int d^3x \bar{\gamma}^{ij} \mathfrak{t}_{ij}\right) \nonumber \\
+\frac{\kappa}{2}  \int d^3 x d^3x' \left( \frac{\mathfrak{e}(x)
[\mathfrak{e}(x') - p(x') -\frac{1}{2} g(x)]}{2 \pi |{\bf x} - {\bf x'}|}  - \mathfrak{p}^i(x) \mathfrak{p}^j(x')
\Delta_{ij}(x-x') \right), \label{hfin}
\end{eqnarray}
where
\begin{eqnarray}
\Delta_{ij}(x) = \int \frac{d^3k}{(2 \pi)^3 k^2} e^{-i {\bf k}
\cdot {\bf x}}  \left(\delta_{ij} - \frac{3k_ik_j}{4k^2}
\right),
\end{eqnarray}
and we wrote  $p(x) = \frac{1}{3} \partial_i \phi \partial^i \phi$ and $g(x) = m^2 \phi^2$.

\section{Master Equation for Gravitational Decoherence of Quantum Matter}

In this section, we quantize the system described in Sec. 2. We trace out the gravitational degrees of freedom and derive a master equation for the quantum matter field. Then, we project the system to the one-particle subspace, in order to derive the evolution equations for a single particle. The resulting master equation simplifies significantly in the non-relativistic regime. We solve the master equation in this regime and we identify the decoherence time.

\subsection{The Hamiltonian operator}

We next proceed to the quantization of the physical degrees of
freedom appearing in the Hamiltonian Eq. (\ref{hfin}).
We write the quantum operator representing the free part $\int d^3 x \mathfrak{e}$
of the Hamiltonian as $\hat{H}_0$ and the operator representing the
gravitational self-interaction as $\kappa \hat{V}_{g}$. Both
operators act on the matter degrees of freedom. At the moment, we do
not specify their exact form, because we want to write a master
equation valid for a general matter content.

We express the quantized transverse-traceless perturbations in terms
of creation and annihilation operators (we drop the tildes on them
henceforth)
\begin{eqnarray}
\hat{h}_{ij}(x) = \sqrt{\frac{2}{\kappa}} \sum_r  \int \frac{d^3 k}{(2\pi)^3 \sqrt{2 {\omega_{\bf k}}}} L_{ij}^r({\bf k})  \left(\hat{b}_r({\bf k}) e^{i {\bf k} \cdot {\bf x}} + b^{\dagger}_r ({\bf k}) e^{-i{\bf k}\cdot{\bf x}}\right),\\ \label{hijq}
\hat{\Pi}_{ij}(x) = -i \sqrt{\frac{\kappa}{2}}  \sum_{r} \int
\frac{d^3 k}{(2\pi)^3 } \sqrt{\frac{\omega_{\bf k}}{2}} L_{ij}^r({\bf
k}) \left(\hat{b}_r({\bf k}) e^{i {\bf k} \cdot {\bf x}} -
b^{\dagger}_r ({\bf k}) e^{-i{\bf k}\cdot{\bf x}}\right), \label{pijq}
\end{eqnarray}
where $r = 1, 2$ denotes the two polarizations, and $\omega_{\bf k} =
\sqrt{k_ik^i}$. The matrices $L_{ij}^r$ are transverse-traceless, and
normalized to satisfy the conditions $\sum_r L_{ij}^r({\bf k})
L_{kl}^r({\bf k}) = \frac{1}{2}(P_{ik}P_{jl} + P_{il}P_{jk})$, where
$P_{ij} = \delta_{ij} - k_ik_j/k^2$ is the projector onto the
transverse direction.

The operator representing the Hamiltonian Eq. (\ref{hfin}) is
\begin{eqnarray}
\hat{H} = \hat{H}_0 + \kappa \hat{V} + \sum_r \int\frac{d^3k}{(2
\pi)^3} \omega_{\bf k} \hat{b}^{\dagger}_r({\bf k}) \hat{b}_r({\bf
k}) \nonumber \\
 - \sqrt{\frac{\kappa}{2}} \sum_r \int \frac{d^3k}{(2\pi)^3\sqrt{2
\omega_{\bf k}}} \left[\hat{b}_r({\bf k} ) \hat{J}^{\dagger}_r({\bf
k}) + b^{\dagger}_r({\bf k}) \hat{J}_r({\bf k})\right], \label{hq}
\end{eqnarray}
where $[\hat{b}_r({\bf k}),\hat{b}_s({\bf k'})] = [\hat{b}^{\dagger}_r({\bf k}),\hat{b}_s({\bf k'})] =0$, $[\hat{b}_r({\bf k}),\hat{b}^{\dagger}_s({\bf k'})] = \delta({\bf k} - {\bf k'}) \delta_{rs}$. We  defined the operators $\hat{J}_r({\bf k}) = \hat{J}^{\dagger}_r(-{\bf k})$ as`
\begin{eqnarray}
\hat{J}_r({\bf k})= L^r_{ij}({\bf k})
 \int d^3x e^{-i {\bf k} \cdot{\bf x}} \hat{\mathfrak{t}}^{ij}(x), \label{jr}
\end{eqnarray}
where $\hat{\mathfrak{t}}^{ij}(x)$ is the (normal-ordered) quantum operator
representing the stress-tensor  in Eq. (\ref{hfin}).

Eq. (\ref{hq}) shows that the environment consists of  a collection
of harmonic oscillators coupled to the matter degrees of freedom. The coupling is linear with respect to the creation and annihilation operators of the environment oscillators. The system is formally similar to a quantum Brownian motion (QBM) model, with the transverse traceless degrees of freedom playing the role of the bath oscillators. In order to compare with the standard QBM models, we note that for the system consisting of a single-particle---a case considered in Refs. \cite{An96, HaKl}--- the interaction Hamiltonian between system and environment is proportional to $\hat{p}^2 \hat{q}_i$, where $\hat{q}_i$ is the position operator of the environment oscillators (the gravitational perturbations) and $\hat{p}^2$ is the particle's momentum.

Another important point is that the Hamiltonian Eq. (\ref{hq}) follows from the  expansion of the gravitational action Eq.(\ref{S3+1}) in powers of $\kappa$, keeping only the quadratic terms in the perturbations. This expansion leads to the standard formulation of linearized perturbations in general relativity. However,   for microscopic manifestations of gravity, beyond the usual applications of general relativity, one might devise non-standard gravitational expansion schemes with whatever rationale. These could lead to open-system dynamics that differ significantly from QBM models.

\subsection{The initial state of the gravitational field}

Next, we trace out the gravitational part of the action. To this end, we assume that the initial state of
the combined system factorizes as $\hat{\rho}_{mat} \otimes \hat{\rho}_{B}$. The specification of the state $\hat{\rho}_B$ of the gravitational perturbations will determine the physics described by the master equation. There are two possibilities.
\begin{enumerate}{}
\item The usual ideas about the quantization of the gravitational field suggests that Minkowski spacetime is the ground state of a quantum gravity theory and that the perturbations are inherently quantum. In this line of thought the quantum fields $\hat{h}_{ij}(x)$ represent gravitons, and the natural choice of $\hat{\rho}_B$ is the graviton's vacuum. At finite temperatures the behavior of gravitons  has been treated in \cite{CamHu,APV}. Note however that gravitons interact very weakly and thus their thermalization cannot be automatically assumed as in baths made of other more strongly interacting particles.
\item Alternatively, one may consider that the spacetime description is already classical, arising as a thermodynamic/hydrodynamic limit of an underlying theory. In this emergent viewpoint, the perturbations around the Minkowski spacetime are  {\em classical}, and their fluctuations  stochastic, rather than quantum. In this perspective, the state $\hat{\rho}_B$ should be  classical, in the sense that its correlation functions  correspond to the thermodynamic/statistical fluctuations of a classical effective
field.
\end{enumerate}

The two possibilities above differ in whether they view the Minkowski
spacetime as the lowest energy microstate, or the lowest energy
macrostate of the collective variables derived from an underlying
theory of quantum gravity.

We want to choose a state $\hat{\rho}_B$ that interpolates between
the two alternatives. The state should  be stationary, reflecting the
time-translation symmetry of Minkowski spacetime. Assuming that it is
also a Gaussian state, the only choice is a thermal state at a
"temperature" $\Theta$. According to the discussion above, $\Theta$
{\em should not} be viewed as a temperature of the graviton
environment, but as a phenomenological parameter interpolating
between the fully quantum and the classical/stochastic regime of
gravitational fluctuations. An analogue of this interpretation for $\Theta$ is that of {\em noise temperature} \cite{comsys}, i.e.,   a parameter characterizing the power spectral density of the noise in stochastic systems, that is not related to a thermodynamic temperature.

In fact, the specific form of the initial state may not affect significantly the physical predictions in certain regimes. For a single non-relativistic particle (Sec. 3.5), only the behavior of the state in the deep infrared sector of the environment ($\omega \rightarrow 0$) contributes to the non-unitary dynamics.

A second parameter characterizing the environment is a cut-off scale $\Lambda$ for the energy of the
perturbations. In order to avoid particle creation effects, it is necessary to assume that $\Lambda$ is much smaller than the masses of any particles in the theory.


A potential problem in the choice of the  state above for the gravitational perturbations is that a thermal state is not Lorentz invariant. The ensuing open system dynamics would then lead to a breaking of Lorentz invariance of the field. However, in the present context, Lorentz invariance has been broken by gauge-fixing prior to quantization. There is no physical representation of the Lorentz group in the Hilbert space of the quantized gravitational perturbations, so we do not know {\em a priori} the rule under which a thermal state transforms with the changes of coordinate systems. We can {\em postulate} that the chosen initial state remains unchanged when transforming from one frame to another, or, more plausibly, that the thermal state is an approximation to a state that remains invariant under the, yet unknown, physical representation of the Lorentz group. In this perspective, the correct rule for Lorentz transformations can be obtained only if we have a gauge-invariant prescription for quantization of the matter-gravity system.

 \subsection{The second-order master equation for the matter fields}

Tracing out the gravitational degrees of freedom yields to second-order in $\sqrt{\kappa}$ the master equation for the reduced density matrix $\hat{\rho}_t$ of the matter fields \cite{Dav, BP}
\begin{eqnarray}
\frac{\partial \hat{\rho}_t}{\partial t} = -i[\hat{H}_0 + \frac{\kappa}{2} \hat{V},\hat{\rho}_t]
\nonumber
\\
-  \frac{\kappa}{4} \sum_a \frac{\coth \left( \frac{\omega_a}{2\Theta}\right)}{\omega_a} \left([\hat{J}_a^{\dagger},[\hat{\tilde{J}}_a(\omega_a), \hat{\rho}_t]] + [\hat{J}_a,[\hat{\tilde{J}}_a^{\dagger}(\omega_a),\hat{\rho}_t]] \right) \nonumber \\
- \frac{\kappa}{4} \sum_a \frac{1}{\omega_a} \left(\{\hat{J}_a^{\dagger},[\hat{\tilde{J}}_a(\omega_a), \hat{\rho}_t]\} - \{\hat{J}_a,[\hat{\tilde{J}}^{\dagger}_a(\omega_a), \hat{\rho}_t]\} \right), \label{master1}
\end{eqnarray}
where we used the combined index $a$ to denote the pair $({\bf k},
r)$ such that $\sum_r \int \frac{d^3k}{(2\pi)^3} \rightarrow \sum_a$,
$\hat{J}_r({\bf k}) \rightarrow \hat{J}_a$ and so on. The operator
$\hat{\tilde{J}}$ is defined as
\begin{eqnarray}
\hat{\tilde{J}}_a(\omega) = \int_0^{\infty} ds e^{-i \omega s} e^{-i \hat{H}_0s} \hat{J}_a e^{i \hat{H}_0s}. \label{tjr}
\end{eqnarray}

The master equation (\ref{master1}) has constant coefficients and it is of the Lindblad type \cite{Lind}. Its derivation (to second order in $\sqrt{\kappa}$ does not require the Born and the Markov approximation, only the condition that the coupling is very small \cite{Dav}. It does not hold for times much larger than the relaxation time, but this is not a problem for the study of decoherence.

Eq. (\ref{master1}) is expressed in a form that applies to any matter field described by the Hamiltonian Eq. (\ref{hfull}). It can be used to describe the effect of gravitational perturbations on the electromagnetic or any other quantum field.

In this paper, we focus on the case of spinless particles of mass $m$ described by a real scalar field $\hat{\phi}(x)$. We express the operators $\hat{\tilde{J}}$ in the master equation in terms of the field's creation and annihilation operators. Then, we obtain the master equation for a single particle, by restricting the field density matrix into the one-particle subspace.

The reason we treat the matter degrees of freedom as a quantum field and then restrict to the one-particle subspace for the description of particle motion, is that doing it this way avoids any ambiguity associated with the choice of coupling between the particles and the gravitational field. The minimal coupling between the matter field and gravity expressed through the Laplace-Beltrami operator in curved background spacetime (\ref {Scov}) is a local term in the Hamiltonian.  In contrast, in a treatment that starts from particles coupled to the gravitational field, the interaction term would be of the form
\begin{eqnarray}
\hat{H}_{int} = \int d^3 x f({\bf x} - {\bf q} ) \hat{A}_{ij} \hat{h}^{ij}({\bf x}),
\end{eqnarray}
where $A_{ij}$ is an operator on the particle's Hilbert space, ${\bf q}$ represents the position of the particle, and $f$ is a phenomenological function that needs to be inserted in order to describe the localization of the interaction, taking into account the finite dimensions of the particle. There is no fixed rule that allows for the determination of the function $f$ from first principles, as is necessary in a treatment of gravitational decoherence. A quantum field theory treatment of particle-field interaction is more fundamental and avoids the ambiguities in the choice of  couplings.

Proceeding to the computation of the operators $\hat{J}_r({\bf k})$,  we decompose the quantum operator $\hat{\phi}(x)$ for the field in terms of creation and annihilation operators
\begin{eqnarray}
\hat{\phi}(x) = \int \frac{d^3 p}{(2 \pi)^3\sqrt{2 \omega_{\bf p}}}
\left( \hat{a}_{\bf p} e^{i {\bf p}\cdot {\bf x}} +
\hat{a}^{\dagger}_{\bf p} e^{- i {\bf p}\cdot{\bf x}} \right),
\end{eqnarray}
where $[\hat{a}_{\bf p}, \hat{a}_{\bf p'}] = [\hat{a}^{\dagger}_{\bf p},\hat{a}^{\dagger}_{\bf p'}] = 0 $, $[\hat{a}_{\bf p}, \hat{a}^{\dagger}_{\bf p'}] = \delta_{\bf p p'}$, and $\omega_{\bf p} = \sqrt{{\bf p}^2 + m^2}$.

From Eq. (\ref{jr}), we  obtain
\begin{eqnarray}
\hat{J}_{r}({\bf k}) =  L_{ij}^r({\bf k}) \int \frac{d^3p}{(2 \pi)^3}
\frac{p^i p^j}{ \sqrt{2\omega_{\bf p}}} \left(\frac{\hat{a}_{\bf p}
\hat{a}_{\bf k-p}}{\sqrt{2 \omega_{\bf k - p}}} +
\frac{\hat{a}^{\dagger}_{\bf p} \hat{a}^{\dagger}_{\bf -k-p}}{\sqrt{2
\omega_{\bf k + p}}} +  2 \frac{\hat{a}^{\dagger}_{\bf p}
\hat{a}_{\bf k+ p}}{\sqrt{2 \omega_{\bf k + p}}}\right). \label{jr2}
\end{eqnarray}

In the derivation of Eq. (\ref{jr2}) we have used the normal ordered
form of the operator $\hat{t}_{ij}(x) = \partial_i\hat{\phi}(x)
\partial_j \hat{\phi}(x)$.

In order to compute the operator $\hat{\tilde{J}}_a(\omega)$ we write
\begin{eqnarray}
f(\omega) = \int_0^{\infty} ds e^{- i \omega s} = \pi \delta (\omega)
- i PV(\frac{1}{\omega}),
\end{eqnarray}
where $PV$ denotes the Cauchy principal part. When evaluating
$\hat{\tilde{J}}_a(\omega)$ according to Eq. (\ref{tjr}), the terms
in Eq. (\ref{jr2}) involving two creation or two annihilation
operators are multiplied by $f(\omega_{\bf p} +\omega_{\bf p'} \pm
\omega)$. Since $\omega_{\bf p} >> m$, and the frequencies of the
environment are bounded by a cut-off  $\Lambda << m$, their
contribution is suppressed in comparison to the other terms, which
are multiplied by $f(\omega_{\bf p} - \omega_{\bf p'} \pm \omega)$.
Hence,
\begin{eqnarray}
\hat{\tilde{J}}_r({\bf k}, \omega) \simeq L_{ij}^r({\bf k}) \int
\frac{d^3p}{(2 \pi)^3} \frac{p^i p^j}{ \sqrt{2\omega_{\bf p}}}
\frac{\hat{a}^{\dagger}_{\bf p} \hat{a}_{\bf k+ p}}{\sqrt{\omega_{\bf
k + p}}}              f(\omega_{\bf k +p} - \omega_{\bf p} + \omega).
\label{jtilde}
\end{eqnarray}

The term $\hat{V}$ describing  gravitational self-interaction is
\begin{eqnarray}
\hat{V} = - \int \frac{d^3k}{(2\pi)^3}
\left(\frac{2\hat{\mathfrak{e}}^{\dagger}_{\bf k} \mathfrak{e}_{\bf
k}}{k^2} - \frac{2\hat{\mathfrak{e}}_{\bf k}^{\dagger}\hat{p}_{\bf
k}}{k^2} + (\delta_{ij} - \frac{3k_i k_j}{4k^2})
\frac{\hat{\mathfrak{p}}^{i\dagger}_{\bf k} \hat{\mathfrak{p}}^j_{\bf
k}}{k^2} - \frac{\hat{\mathfrak{e}}^{\dagger}_{\bf k} \hat{g}_k}{k^2}\right), \label{V},
\end{eqnarray}
expressed in terms of the normal-ordered operators
\begin{eqnarray}
\hat{\mathfrak{e}}_{\bf k} &=& \int \frac{d^3p}{(2 \pi)^3} \left( \frac{ \omega_{\bf p}^2 -\omega_{\bf p} \omega_{\bf k+p} + {\bf p}\cdot{\bf k}}{4\sqrt{\omega_{\bf p} \omega_{\bf k+p} }} \hat{a}_{\bf p} \hat{a}_{\bf -p-k} + \frac{\omega_{\bf p}^2 -\omega_{\bf p} \omega_{\bf p-k}  - {\bf p}\cdot{\bf k}}{4\sqrt{\omega_{\bf p} \omega_{\bf p-k} }} \hat{a}^{\dagger}_{\bf p} \hat{a}^{\dagger}_{\bf k-p}\right. \nonumber \\
 &+& \left. \frac{\omega_{\bf p}^2 + \omega_{\bf p} \omega_{\bf p-k} - {\bf p}\cdot{\bf k}}{2\sqrt{\omega_{\bf p} \omega_{\bf p-k} }} \hat{a}^{\dagger}_{\bf p} \hat{a}_{\bf p-k} \right)
\\
\hat{p}_{\bf k} &=& \frac{1}{6}\int \frac{d^3p}{(2\pi)^3}
\left(\frac{{\bf p}^2 + {\bf p}\cdot {\bf k}}{\sqrt{\omega_{\bf p}
\omega_{p+k}}} \hat{a}_{\bf p} \hat{a}_{\bf -p-k}
 \frac{{\bf p}^2 -
{\bf p}\cdot {\bf k}}{\sqrt{\omega_{\bf p} \omega_{p-k}}}
\hat{a}^{\dagger}_{\bf p} \hat{a}^\dagger_{\bf k-p}
\nonumber \right. \\
&+& \left.
 + 2 \frac{{\bf
p}^2 - {\bf p}\cdot {\bf k}}{\sqrt{\omega_{\bf p} \omega_{p-k}}}
\hat{a}^{\dagger}_{\bf p} \hat{a}_{\bf p-k} \right)
\\
\hat{\mathfrak{p}}^i_{\bf k} &=& \frac{i}{2} \int
\frac{d^3p}{(2\pi)^3} \left[ \sqrt{\frac{\omega_{\bf p}}{\omega_{\bf
p+k}}} (p^i+k^i) \hat{a}_{\bf p} \hat{a}_{\bf -k-p} +
\sqrt{\frac{\omega_{\bf p}}{\omega_{\bf p-k}}} (p^i - k^i)
\hat{a}_{\bf p} \hat{a}_{\bf k-p} \right.
\nonumber \\
 &+& \left. \left(\sqrt{\frac{\omega_{\bf p}}{\omega_{\bf p-k}}}  (p^i-k^i) + \sqrt{\frac{\omega_{\bf p-k}}{\omega_{\bf p}}}p^i\right) \hat{a}^{\dagger}_{\bf p}\hat{a}_{\bf p-k}\right]
 \\
 \hat{g}_{\bf k} &=& m^2 \int
\frac{d^3p}{(2\pi)^3} \left[ \frac{1}{2 \sqrt{\omega_{\bf p} \omega_{\bf k+p} }}  \hat{a}_{\bf p} \hat{a}_{\bf -p-k} + \frac{1}{2 \sqrt{\omega_{\bf p} \omega_{\bf p-k} }} \hat{a}^{\dagger}_{\bf p} \hat{a}^{\dagger}_{\bf k-p}
\nonumber \right. \\
&+& \left.
+ \frac{1}{ \sqrt{\omega_{\bf p} \omega_{\bf p-k} }} \hat{a}^{\dagger}_{\bf p} \hat{a}_{\bf p-k}\right].
\end{eqnarray}

\subsection{Restriction to the one-particle subspace}

Since we want to write a master equation describing the evolution of
a single particle, we restrict the density matrix $\hat{\rho}$ into
the single-particle subspace ${\bf H}_1$ of the Hilbert space ${\cal
H}$ of the field \cite{An97,AnZ}. A single-particle state is expressed in the field
Hilbert space as $\int \frac{d^3p}{(2 \pi)^3} \psi({\bf p})
\hat{a}^{\dagger}_{\bf p}|0\rangle$, where $\psi({\bf p})$ is the
particle's wave-function in momentum space and $|0\rangle$ is the
field vacuum. The density matrix for a single particle $\hat{\rho}_1$
is thus represented by the field density matrix
\begin{eqnarray}
\hat{\rho} = \int \frac{d^3p}{(2\pi)^3}\frac{d^3p'}{(2\pi)^3}
\rho_1({\bf p},{\bf p'}) \hat{a}^{\dagger}_{\bf p} |0\rangle \langle
0| \hat{a}_{\bf p'}, \label{ro1}
\end{eqnarray}
where $\rho_1({\bf p},{\bf p'}) = \langle {\bf p}|\hat{\rho}_1|{\bf p'}\rangle_{{\cal H}_1}$ is the single-particle density matrix in the momentum representation.

We then project the master equation (\ref{master1}) to ${\cal H}_1$. To this end, we substitute a density matrix of the form Eq. (\ref{ro1}) into Eq. (\ref{master1}) and retain only the terms that preserve this form.

The von Neumann term $-i [\hat{H}_0, \hat{\rho}]$ for the free
Hamiltonian preserves the single-particle subspace, giving rise to a
term $-i [\sqrt{\hat{{\bf p}}^2 + m^2}, \hat{\rho}_1]$ on ${\cal
H}_1$ representing the evolution of a free relativistic particle.

\paragraph{The non-unitary terms.} To project the non-unitary terms of Eq. (\ref{master1}) into ${\cal H}_1$ we proceed as follows. The commutators with the $\hat{\tilde{J}}_a$ operators of Eq. (\ref{jtilde}) preserve the single-particle subspace. The only terms that fail to preserve ${\cal H}_1$ are the components of $\hat{J}_a$ involving two creation or two annihilation operators in Eq. (\ref{jr2}). Dropping these terms we find that the projection of the non-unitary terms correspond to a superoperator ${\bf L}$ on ${{\cal H}_1}$ defined by

\begin{eqnarray}
{\bf L}[\hat{\rho}_1] = -\frac{\kappa}{4} \sum_a \left[\frac{\coth \left( \frac{\omega_a}{2\Theta}\right)}{\omega_a} \left([\hat{A}_a^{\dagger},[\hat{B}_a,\hat{\rho}_1]] + [\hat{A}_a,[\hat{B}^{\dagger}_a, \hat{\rho}_1]]\right)
\right.
\nonumber \\
\left. - \frac{1}{\omega_a} \left( \{ \hat{A}^{\dagger}_a,[\hat{B}_a, \hat{\rho}_1]\} - \{\hat{A}_a,[\hat{B}^{\dagger}_a, \hat{\rho}_1]\} \right) \right],
\end{eqnarray}
where $\hat{A}_a \equiv \hat{A}_r({\bf k})$ and $\hat{B}_a \equiv \hat{B}_r({\bf k})$ are operators on ${\cal H}_1$ defined by their matrix elements in the momentum basis
\begin{eqnarray}
\langle {\bf p}| \hat{A}_r({\bf k})|{\bf p'}\rangle = L_{ij}^r({\bf k}) \frac{p^ip^j}{\sqrt{\omega_{\bf p} \omega_{{\bf p'}}}} (2 \pi)^3 \delta({\bf p'}- {\bf p} - {\bf k}) \label{ar}\\
\langle {\bf p}| \hat{B}_r({\bf k})|{\bf p'}\rangle = \langle {\bf p}| \hat{A}_r({\bf k})|{\bf p'}\rangle f(\omega_{\bf p} - \omega_{\bf p'} + \omega_{\bf k}).
\end{eqnarray}

\paragraph{The gravitational self-interaction.} The projection of the von Neumann term describing gravitational self-interaction onto ${\cal H}_1$ yields a term $-i\frac{ \kappa}{2} [\hat{U}, \hat{\rho}_1]$, where
\begin{eqnarray}
\hat{U} = -\int \frac{d^3k}{(2 \pi)^3} \frac{1}{k^2} \left[F^1_{\bf
k}(\hat{{\bf p}}) + F^2_{\bf k}(\hat{{\bf p}}) + F^3_{\bf
k}(\hat{{\bf p}}+ F^4_{\bf
k}(\hat{{\bf p}}) \right].
\end{eqnarray}
The operators $F^i_{\bf k}(\hat {\bf p})$ are functions of the
3-momentum operator $\hat{p}^i$ for a relativistic particle. Each corresponds to one of the terms in the sum of Eq. (\ref{V}), as they are projected in the single-particle Hilbert space. In particular, $F^1_{\bf k} $ corresponds to the term $2\hat{\mathfrak{e}}^{\dagger}_{\bf k} \mathfrak{e}_{\bf
k}$, $F^2_{\bf k}$ corresponds to $- 2\hat{\mathfrak{e}}_{\bf k}^{\dagger}\hat{p}_{\bf
k}$, $F^3_{\bf k}$ corresponds to
$ (\delta_{ij} - \frac{k_i k_j}{2k^2}) \hat{\mathfrak{p}}^{i\dagger}_{\bf k} \hat{\mathfrak{p}}^j_{\bf
k}$, and $F^4_{\bf k}$ corresponds to $- \hat{\mathfrak{e}}^{\dagger}_{\bf k} \hat{g}_k$. Their explicit form is the following
\begin{eqnarray}
F^1_{\bf k}(\hat{\bf p}) &=& 2\omega_{\bf p} \omega_{{\bf p}+{\bf k}} - \frac{\omega_{\bf p}}{ \omega_{{\bf p}+{\bf k}}} ({\bf p}\cdot {\bf k} + {\bf k}^2) -  {\bf p} \cdot{\bf k},
\\
F^2_{\bf k}(\hat{\bf p}) &=& -\frac{ {\bf p} \cdot ({\bf p} + {\bf k})}{3 \omega_{\bf p}\omega_{{\bf p}+{\bf k}}} (2 \omega_{\bf p}^2 -{\bf k}^2)
\\
F^3_{\bf k} (\hat{\bf p}) &=& -\frac{1}{4} \left[ \frac{\omega_{{\bf p}+{\bf k}}}{\omega_{\bf p}} \left({\bf p}^2 -\frac{3 ({\bf p} \cdot{\bf k})^2}{4 k^2}\right) + \frac{\omega_{\bf p} }{\omega_{{\bf p}+{\bf k}}} \left( ({\bf p} + {\bf k})^2 - \frac{ 3[({\bf p}+{\bf k})\cdot {\bf k}]^2}{4k^2}\right) \right. \nonumber \\
  &+& \left. 2 {\bf p}^2 + \frac{1}{2}{\bf k} \cdot {\bf p} - \frac{3({\bf p} \cdot{\bf k})^2}{2k^2} \right]\\
  F^4_{\bf k} &=& -\frac{m^2}{4} \left( \frac{3 \omega_{\bf p}}{\omega_{\bf p+k}} + \frac{\omega_{\bf p+k}}{\omega_{\bf p}} + 2 {\bf p} \cdot {\bf k} - {\bf k}^2 \right).
\end{eqnarray}

Thus, the master equation for a single relativistic particle is
\begin{eqnarray}
\frac{\partial \hat{\rho}_1}{\partial t} = - i [\sqrt{m^2 + \hat{{\bf
p}}^2}, \hat{\rho}_1] - i \frac{\kappa}{2} [\hat{U}, \hat{\rho}_1] + {\bf
L}[\hat{\rho}_1]. \label{mrel}
\end{eqnarray}

\subsection{The non-relativistic limit}

The master equation (\ref{mrel}) is still very complex. However, it simplifies significantly in the non-relativistic limit.
For $|{\bf p}| <<m$, the matrix elements of the operators ${\bf A}_a$ become
\begin{eqnarray}
\langle {\bf p}| \hat{A}_r({\bf k})|{\bf p'}\rangle \simeq
L_{ij}^r({\bf k}) \frac{p^ip^j}{m} (2 \pi)^3 \delta({\bf p'}- {\bf p}
- {\bf k}),
\end{eqnarray}
and thus $\hat{A}_r$ can be expressed as
\begin{eqnarray}
\hat{A}_r({\bf k}) = L^r_{ij}({\bf k}) \frac{\hat{p}^i\hat{p}^j}{m}
e^{i k_i \hat{X}^i}, \label{eqA}
\end{eqnarray}
where $\hat{x}^i$ is the position and $\hat{p}_j$ the momentum
operators of a non-relativistic particle.

In order to calculate $\hat{B}_a$ at the non-relativistic limit, we note that $\omega_{\bf p} - \omega_{\bf p+k} + \omega_{\bf k} \simeq \omega_{\bf k} - \frac{1}{2} \omega_{\bf k}^2/m - {\bf k}\cdot{\bf p}/m \simeq \omega_{\bf k}$, where the last step follows because $|{\bf p}|<<m$ and $\omega_{\bf k} << m$. Thus, we obtain
\begin{eqnarray}
\hat{B}_r({\bf k}) = f(\omega_{\bf k}) \hat{A}_r({\bf k}).
\end{eqnarray}
The non-unitary part of the master equation then becomes
\begin{eqnarray}
{\bf L}[\hat{\rho}_1] &=& -\frac{\kappa}{4} \sum_a \pi
\delta(\omega_{\bf k}) \left[\frac{\coth \left(
\frac{\omega_a}{2\Theta}\right)}{\omega_a}
\left([\hat{A}_a^{\dagger},[\hat{A}_a,\hat{\rho}_1]] +
[\hat{A}_a,[\hat{A}_a^{\dagger}, \hat{\rho}_1]]\right) \right.
\nonumber \\
&-& \left. \frac{1}{\omega_a} \left( \{ \hat{A}^{\dagger}_a,[\hat{A}_a, \hat{\rho}_1]\} - \{\hat{A}_a,[\hat{A}^{\dagger}_a, \hat{\rho}_1]\} \right) \right] \nonumber \\
&+&\frac{i\kappa}{4} \sum_a PV(1/\omega_{\bf k}) \left[\frac{\coth \left( \frac{\omega_a}{2\Theta}\right)}{\omega_a} \{[\hat{A}_a^{\dagger},\hat{A}_a], \hat{\rho}_1\} -  \frac{1}{\omega_a}  [ \{ \hat{A}^{\dagger}_a, \hat{A}_a\}, \hat{\rho}_1]\right] \hspace{1cm} \label{master3}
\end{eqnarray}
Using Eq. (\ref{eqA}) for $\hat{A}_a$, we evaluate Eq. (\ref{master3}) to obtain
\begin{eqnarray}
{\bf L}[\hat{\rho}_1] =  - \frac{ \kappa \Theta}{18m^2} (\delta^{ij}
\delta^{kl} + \delta^{ik}\delta^{jl}) [\hat{p}_i
\hat{p}_j,[\hat{p}_k\hat{p}_l, \hat{\rho}_1]] - i\frac{ \kappa
\Lambda}{9\pi m^2} [\hat{p}^4, \hat{\rho}_1]. \label{lll}
\end{eqnarray}

To first order in $|{\bf p}|^2/m^2$, the functions $F^i_{\bf k}$ appearing in the self-interaction term $\hat{U}$ become
\begin{eqnarray}
F^1_{\bf k}({\bf p}) \simeq 2 m^2 + 2 {\bf p}^2 \\
F^2_{\bf k}({\bf p}) \simeq -\frac{2}{3} ({\bf p}^2 +{\bf p}\cdot {\bf k})\\
F^3_{\bf k}({\bf p}) \simeq - {\bf p}^2 -\frac{{\bf k}^2}{16} + {\bf p} \cdot{\bf k} + \frac{3({\bf p} \cdot{\bf k})^2}{4k^2},\\
F^4_{\bf k}({\bf p}) \simeq - m^2 + \frac{1}{2}{\bf k}^2
\end{eqnarray}
where we ignored terms of order $\Lambda^2/m^2$ since the cut-off has been assumed to be much smaller than the particle's rest mass.
Hence, the operator $\hat{U}$ becomes
\begin{eqnarray}
\hat{U} =  - \frac{ \Lambda m^2}{ 2\pi^2}  - \frac{ 7\Lambda}{ 24\pi^2} {\bf p}^2
\label{U}
\end{eqnarray}

 The first term in Eq. (\ref{U}) is a constant, and thus, it drops from the master equation. The second term is proportional to the free-particles energy and thus it corresponds to a renormalization of the mass. The renormalized mass is

\begin{eqnarray}
m_{R} = m (1 + \frac{ 7 \kappa \Lambda m}{24 \pi^2}). \label{massrel}
\end{eqnarray}
Note, that the mass renormalization Eq. (\ref{massrel}) applies only to second order in perturbation theory. The full theory describing the interaction of gravity with a scalar field, as is well-known, is not renormalizable. There is no contradiction in the effective field theory sense because in the present context we are interested only in the low energy, weak gravity dynamics.

The unitary term  involving ${\bf p}^4$ in Eq. (\ref{lll}) contributes a correction to higher-order kinetic-energy terms. Thus, it may be ignored when considering only the non-relativistic expression for the particle's energy.

 We have thus obtained  the master equation for a non-relativistic particle interacting with gravity, valid to first order in $\kappa$.
\begin{eqnarray}
\frac{\partial \hat{\rho}_1}{\partial t} = - \frac{i}{2m_R}
[\hat{{\bf p}}^2,\hat{\rho}_1]- \frac{ \kappa \Theta}{18m_R^2}
(\delta^{ij} \delta^{kl} + \delta^{ik}\delta^{jl}) [\hat{p}_i
\hat{p}_j,[\hat{p}_k\hat{p}_l, \hat{\rho}_1]] \label{me1}
\end{eqnarray}

\subsection{Decoherence of a Quantum Particle in a Gravitational Field}

Next, we specialize to the case of motion in one spatial dimension. Then, the master equation becomes

\begin{eqnarray}
 \frac{\partial \hat{\rho}}{\partial t}= - \frac{i}{2m_R} [\hat{p}^2,\hat{\rho}]- \frac{ 4 \pi G \Theta}{9m_R^2}  [\hat{p}^2 ,[\hat{p}^2, \hat{\rho}]]. \label{me1c}
\end{eqnarray}
where we reinserted Newton's constant by setting $\kappa = 8 \pi G$.

This master equation is exactly solvable in the momentum representation
\begin{eqnarray}
\rho_t(p,p') = \exp \left[ - \frac{i}{2m_R} (p^2 - p'^2)t - \frac{ 4
\pi G \Theta}{9m_R^2} (p^2-p'^2)^2 t \right]\rho_0(p,p')
\end{eqnarray}

It is evident that the master equation leads to decoherence in the
energy basis. This is to be expected, because in the non-relativistic limit, a particle couples to the transverse-traceless perturbations through the square of its momentum $p$.

Let us assume that the initial state is a
superposition of two states, localized in momentum $p_1$ and $p_2$.
Define the mean momentum $p = (p_1 +p_2)/2$ and $\Delta p = |p_2 -
p_1|$. Then, after time of order of
\begin{eqnarray}
t_{dec} = \frac{m_R^2}{ G \Theta p^2 \Delta p^2} = \frac{1}{G \Theta m_R^2 v^2
\delta v^2},
\end{eqnarray}
the momentum superpositions will have been destroyed; $v$ and $\delta v$ refer to the mean velocity and the velocity difference, respectively. Inserting back $c$ and $\hbar$ the decoherence time is
\begin{eqnarray}
t_{dec} = \frac{\hbar^2 c^5}{G \Theta m_R^2 v^2 \delta v^2}.
\end{eqnarray}
Note that the decoherence being in the energy basis, rather than in the momentum basis, e.g., the interference between two states peaked at $p$ and $-p$ is not destroyed  \footnote{We thank one referee for alerting us to emphasize this difference.}.

Note the gravitational decoherence time depends not only on the Planck scale, but also on an additional parameter, here, $\Theta$, whose meaning we have explained earlier. This is a generic feature in the open system approach taken here. 
Note also \textit{the Newtonian force term}, that involves only Newton's constant,  always appears in the Hamiltonian part of the evolution equation and it \textit{does not lead to decoherence}. Decoherence is due to the transverse-traceless (TT) perturbations and the corresponding non-unitary term will involve parameters corresponding to the unequal-time correlation function that characterizes the perturbations. These parameters are, in principle, determined by the detailed features of the environment, like the spectral density of a harmonic oscillator bath \cite{QBM}. Here with gravity as the environment it refers to the way Minkowski spacetime is formed by its underlying constituents, what we referred to as the 'textures' of spacetime.   Viewed in this perspective, measurements of the gravitational decoherence rate could provide important information about the nature of gravity -- whether it is fundamental or emergent, and may reveal features in the micro-structure of spacetime.

It is important to note that the non-unitary terms in the master equation (\ref{me1}) do not depend significantly on the details of the chosen initial state for the gravitational perturbations. To see this, note that in Eq. (\ref{master3}), $\Theta$ appears in conjunction with terms that blow up as the frequencies $\omega \rightarrow 0$. In effect, the contribution from the very low frequencies in the environment dominates the non-unitary dynamics. This suggests that $\Theta$ is best viewed as a parameter characterizing the strength  of correlations in the deep infrared sector of gravitational perturbations, rather than an effective temperature common to all modes.

For $\Theta = 0$, the non-unitary terms in Eq. (\ref{me1}) vanish. Thus, if Minkowski spacetime corresponds to the ground state of a quantum gravity theory, there is no gravitational decoherence. However, we need to note that Eq. (\ref{me1}) follows from the second-order master equation for open system dynamics, which involves the Markov approximation.   Ref. \cite{An96} derived the master equation for a particle interacting with the transverse-traceless perturbations in the quantum vacuum without the Markov assumption. It was shown that the non-unitary terms indeed vanish at  timescales larger than $\Lambda^{-1}$, but a transient non-unitary term, similar to the one in  Eq. (\ref{me1}), is present at earlier times. This term leads to a suppression of interferences in the energy basis, but this effect is weak in the microscopic and mesoscopic regimes. Furthermore, the decoherence time-scale is of the order of $\Lambda^{-1}$, hence, it is very sensitive to the type of particle-gravity coupling and model specific.

We add that a master equation of the form  Eq. (\ref{me1}) appeared earlier in  \cite{HaKl} from considering the interaction of a non-relativistic particle in the presence of a potential with a bath of thermal gravitons. For decoherence in graviton backgrounds see also \cite{Ha, Rey} but note that controversies exist in the treatment of  the coherence properties of gravitons leading to an over-prediction of the magnitude of graviton's decoherence effects.

\section{Bearings on other gravitational decoherence theories}

With these results we can discuss the bearings on other gravitational decoherence theories. Those with external parameters put in by hand are difficult to compare. But at least the proponents of such alternative theories can now compare their stochastic equations with the present master equation based on QFT + GR,  so one can see clearly how and where they deviate from the established theories. The transformation from a stochastic Schr\"odinger equation to a master equation involves typical procedures for taking single-run trajectories to a reduced density matrix, the reverse direction involves a more challenging unraveling process. There are two types of theories we can compare with. To put it simply, we are in agreement with the original proposal of Penrose \cite{Penrose} but in disagreement with theories which assume time or space fluctuations.  We explain our position in the following.

\subsection{A more formal characterization of Penrose's concern}


Penrose has argued that there exists a conceptual contradiction in the way time is treated in quantum theory and in general relativity \cite{Penrose}---see also \cite{Chr2}. In quantum theory, time is an external parameter manifested in the evolution of the quantum states through Schr\"odinger's equation. In general relativity, time is part of the spacetime structure, which is dynamical and depends on the configuration of the matter degrees of freedom. This contradiction is manifested clearly when considering superposition of macroscopically distinct states for the particles. Each component of the superposition generates a different spacetime. Since there is no canonical way of relating time parameters in different spacetime manifolds, there is a fundamental ambiguity in the choice of the time parameter of the evolution of the quantum state. This ambiguity is manifested even at low energies, when the gravitational interaction can be effectively described by the Newtonian theory. Penrose argues that this ambiguity provides a hint for a physical mechanism of gravity-induced decoherence for superpositions of macroscopically distinct states.

The results of this paper allow us to provide a more precise mathematical characterization of the ambiguity pointed out by Penrose. The ambiguity is related to the implementation of the constraints of general relativity at the quantum level. We recall that the derivation of the master equation (\ref{me1}) required a gauge choice that preserves the Lorentzian foliation.    This gauge choice is standard, and it does not lead to any ambiguities, at least in the linearized theory. It freezes the freedom of time and space  reparameterizations in the theory, so that the matter degrees of freedom are described by time and space coordinates corresponding to a Lorentz frame. This is essential for the quantization of the matter degrees of freedom in terms of ordinary quantum field theory.

However, for any other gauge choice, the matter degrees of freedom are expressed in a non-inertial frame of Minkowski spacetime. They cannot be quantized using the methods of Poincar\'e invariant quantum field theory. In particular, the implementation of time and space reparameterizations in a quantum field theory requires the consideration of unitary transformations of the quantum field between different (non-inertial) foliations \cite{non-inert1}. This is not possible for a spacetime of dimension greater than $2$ \cite{non-inert2}, at least in the context of canonical quantization \cite{hist}.  Thus, it is not possible to compare the quantum theories obtained from different gauge choices and there is no guarantee that the resulting theory is gauge-covariant.

\subsection{Critique on space-time fluctuation-induced decoherence}

Many approaches to gravitational, intrinsic or fundamental decoherence involve the assumption that  decoherence originates from uncertainties or  fluctuations in the specification of time or space coordinates. We will call this class the `spacetime-fluctuation induced' (STFI) decoherence. Some examples of such theories are given in Ref. \cite{IntDec}.  Such an assumption may be made explicitly, or implicitly in the mathematical description of the decoherence effect.  Our results have significant implications for this approach to gravitational decoherence as we now will show.

Typically, models of STFI-decoherence consider gravity-induced fluctuations in the position or temporal co-ordinate of physical events and they model these fluctuations by classical stochastic processes. In Ref. \cite{IntDec}, we argued that the stochastic modeling of temporal fluctuations involves very strong physical assumptions. Temporal fluctuations in a quantum system cannot, in general, be modeled by a stochastic process, unless these fluctuations have been classicalized (by some other agents of decoherence), or they are classical to begin with. Here, we will argue that the assignment of dynamical (stochastic) content to such fluctuations cannot be reconciled with a gravity theory that shares the same fundamental symmetries with general relativity.

Fluctuations in the time or space coordinates of an event are indistinguishable mathematically from time and space reparameterizations of the system, only such reparameterizations are viewed as stochastic. However, as we showed in Sec. 2, time and space reparameterizations are pure gauge variables in classical relativity (in the linearized approximations); they do not have any dynamical content. The invariance of the theory under space and time reparameterizations follows from the diffeomorphism invariance of the classical action, a fundamental symmetry of general relativity.

The assignment of dynamical contents to such fluctuations implies that they are not treated as gauge variables.
Of course, we noted that there is  an ambiguity in the quantum implementation of the symmetry, but this does not imply that space and time reparameterizations are to be viewed as dynamical. 
Doing so violates the fundamental symmetry of classical general relativity.   Moreover, such a modification would have far-reaching implications which goes beyond the gravitationally induced decoherence effects. The diffeomorphism symmetry affects both the dynamics and the kinematics of general relativity, and its abandonment ought to be manifested in other gravitational phenomena.

In fact, it is very easy to construct models of gravitationally-induced decoherence, if we assume that the parameters
 $q_i(x)$ corresponding to space reparameterizations and $\tau(x)$ corresponding to time reparameterizations  fluctuate stochastically. Since general relativity does not provide a guide for the structure of these fluctuations, any choice for the stochastic process that governs the fluctuations of $q_i(x)$ and $\tau(x)$ is {\em ad hoc}.

 The simplest case of such a stochastic process, that leads to a well known master equation, is the following.
  Under a reparameterization in space and time, the spacetime volume element transforms as $dt d^3x \rightarrow dt d^3 x (1 + \dot{\tau} + \partial_i q^i)$. We assume that the histories $\upsilon_t(x) = \dot{\tau}_t(x) + \partial_i q^i_t(x)$  correspond to a Markov process
 \begin{eqnarray}
 \langle \upsilon_t(x)\rangle = 0,
 \langle \upsilon_t(x) \upsilon_t(x') \rangle = \sigma^4 \delta(t-t') \delta^3(x-x'),
 \end{eqnarray}
where $\sigma$ is a constant with dimensions of time.

We then write the time dependent Hamiltonian $H_t(\upsilon) = \int d^3x (1 + \upsilon_t(x)) {\cal H}(x)$, where
${\cal H}(x)$ is the gauge-fixed Hamiltonian density of Eq. (\ref{hfin}).
  For simplicity, we  assume that the transverse-traceless degrees of freedom of the linearized gravitational field are frozen.

 The ensemble-averaged master equation for the matter degrees of freedom is
\begin{eqnarray}
\frac{\partial \hat{\rho}}{\partial t} = -i [\hat{H}, \hat{\rho}] - \frac{\sigma^4}{2} [\hat{H},[\hat{H}, \hat{\rho}]],
\end{eqnarray}
where $\hat{H}$ is the Hamiltonian operator corresponding to $\int d^3x {\cal H}(x)$.

The above master equation is different from the usual one employed in the discussion of gravitational decoherence, because the Newtonian interaction term is contained in the Hamiltonian. Note the parameter $\sigma$ appearing in the master equation has no {\em a priori} relation to the Newton constant. This is a distinct feature of all `STFI' decoherence theories.  Being gauge variables, time and space reparameterizations are decoupled from the Newtonian interaction at the level of the classical theory. Classical general relativity does not provide any information about the strength of such interactions. There is no physical and no mathematical reason to introduce Newton's constant in the definition of the Markovian process. If, indeed, stochastic time and space reparameterizations follow from an underlying physical theory, this theory is so different from General Relativity, that the latter, commonly accepted as providing the best description of classical gravitational effects, does not provide any guide about the underlying stochastic process.  Therefore, despite their names, these theories are not really about gravitational decoherence.


\section{Findings and Implications}

In this paper we gave a first principles derivation of a master equation for matter, our system, described by a quantum field, interacting with an environment of weak gravitational perturbations, treated as a quantum field. We rely on well established theories, namely, quantum field theory and general relativity, making minimal modeling assumptions. We employ the standard methodology of open quantum systems in the tracing out of the gravitational field to derive a master equation for the reduced density matrix of the matter field.

The derivation of the master equation requires the specification of an initial state of gravitational perturbations which depends on the nature of gravity, which could be fundamental (elemental) or effective (emergent). 
We chose to employ a Gaussian state characterized by a phenomenological noise-temperature $\Theta$. This parameter interpolates between the regime of minimal fluctuations ($\Theta = 0$) and the regime of large classicalized perturbations (large values of $\Theta$). In effect, $\Theta$ contains information about the underlying gravitational textures of (in this case, Minkowski) spacetime.

Having derived a master equation for the quantum matter field,  we projected to its single-particle subspace and then took the non-relativistic limit. In this regime, the master equation simplifies significantly, Eq. (\ref{me1}). It is of the Lindblad type \cite{Lind}, with generators corresponding to the square of the particle's momentum. For $\Theta \neq 0$, the master equation leads to decoherence in the energy basis, but to no decoherence in the position basis.

Our findings in this paper have direct implications regarding the structural integrity of many proposals of theories of gravitational decoherence, and broader implications on the constitutional properties of gravity itself, whether it is fundamental or effective.

\begin{enumerate}

\item Our methodology applies, in principle, to any matter configuration compatible with the weak-gravity approximation or for a treatment of gravitational decoherence in photons and neutrinos. It can be straightforwardly generalized to systems of arbitrary $N$-particles, reaching into the mesoscopic domain. It has been suggested \cite{Leggett} that  new physical laws different from quantum mechanics may be at work in this range. Following the approach here one can obtain results of gravitational decoherence in the mesoscopic scale. If these alternative quantum theories can agree to use GR to describe gravity in their predictions, one should be able to see the difference between their predictions and predictions based on GR+QFT, as exemplified in this work.

\item Our master equation is derived from accepted theories for matter and gravity, namely, quantum field theory and general relativity. It is significantly different from other master equations that have been proposed in the literature. This strongly suggests that many proposed theories of gravitational decoherence postulate, implicitly or explicitly,   mechanisms beyond known physics.

\item Time and space reparameterizations, being pure gauge in general relativity, decouple from the terms describing Newtonian interaction, already at the classical level. Hence, Newton's constant does not need to appear in theories of decoherence  due to space and time fluctuations. The proposers of such theories, which, ironically, are not about gravitational decoherence, need to recognize this fact and  bear the burden to explain the physical meaning and origin of whatever new phenomenological parameters they introduce.

\item Gravitational decoherence depends strongly on assumptions about the nature of gravitational perturbations. The usual assumption that Minkowski spacetime is the ground state of quantum gravity would imply that gravitational perturbations are very weak and cannot lead to decoherence. However, if general relativity is a hydrodynamic theory and gravity is in the nature of thermodynamics, Minkowski spacetime should presumably be identified with a macrostate (i.e., a coarse-grained state of the micro-structures). In this case, the perturbations are expected to be much stronger and they may act efficiently as agents of decoherence. Thus, observation of the magnitude and features of gravitational decoherence may reveal the nature of gravity, whether it is elemental or emergent.

\end{enumerate}

\bigskip

{\bf Acknowledgment}    Results from this work were first reported at Peyresq Physics 16 in June, 2011, directed by Prof. Edgard Gunzig, supported by OLAM, Association pour la Recherche Fondamentale, Brussels; again at the Galiano Island meeting on intrinsic decoherence in May 2013, organized by Profs. Philip Stamp and Bill Unruh, supported by the John Templeton Foundation. BLH wants to thank Prof. Ming-Chung Chu of the Chinese University of Hong Kong, Prof. Hsiang-Nan Li of the Institute of Physics, Academia Sinica, Taiwan and Prof. Yong-Shi Wu, Director of the Center for Theoretical Physics, Fudan University, Shanghai, for their kind hospitality during his visits in the Spring of 2013 while this paper was completed.

\bigskip

\bigskip


\end{document}